\documentstyle[12pt]{article}
\def\GeV{{\rm GeV}}
\def\half{{\textstyle{1\over2}}}
\def\roughly#1{\raise.3ex\hbox{$#1$\kern-.75em\lower1ex\hbox{$\sim$}}}
\def\unity{{\rlap{1} \hskip 1.4pt 1}}
\def\br{{\rm BR}}

\begin{document}
\begin{titlepage}
\begin{center}
December 1995\hfill    UND-HEP-95-US01 \\
               \hfill    RU-95-72 \\
               \hfill    hep-ph/9512354
\vskip .2in
{\large \bf
Disoriented Sleptons}
\vskip .3in
Riccardo Rattazzi\footnote{E-mail:
rattazzi@physics.rutgers.edu}\\[.03in]
{\em Department of Physics and Astronomy\\
     Rutgers University\\
     Piscataway, NJ  08855}
\vskip 10pt
Uri Sarid\footnote{E-mail:
sarid@particle.phys.nd.edu}\\[.03in]
{\em Department of Physics\\
     University of Notre Dame\\
     Notre Dame, IN 46556}
\end{center}
\vskip .2in
\begin{abstract}
\medskip
We discuss some fundamental concerns regarding the recent proposal
of Dimopoulos and Giudice for dynamically aligning the soft masses
of the sfermions in the minimal supersymmetric standard model (MSSM)
with
the corresponding fermion masses to suppress flavor changing neutral
currents. We show that the phenomenologically-favored presence of
right-handed neutrinos in the theory, even if only at very high
scales, generically disaligns the slepton mass matrices.
Further suppression is then needed to meet the current upper bound
on the rate for $\mu \to e\gamma$. Planned improvements in the
search for $\mu \to e\gamma$ should easily detect this rare mode.
(With improved sensitivity $\mu\to 3e$ may also be seen.) By
measuring the helicity of the amplitude for $\mu \to e\gamma$ such
experiments could distinguish between unified and non-unified models
at very high energies; by inserting the various MSSM parameters as
they become available, the mixing in the leptonic Yukawa couplings
can be extracted; and by combining the results with those of various
neutrino experiments some information about the right-handed
neutrino Majorana matrices can also be gained.
\end{abstract}
\end{titlepage}

The most promising candidates for a fundamental theory underlying
the standard model have been supersymmetric (SUSY) models. There are
compelling theoretical and phenomenological reasons to believe that
nature is supersymmetric on microscopic scales, and that the
observed asymmetry at low energies between bosons and fermions is
due to spontaneous SUSY breaking. Much attention has been focused on
the highly-successful minimal supersymmetric extension of the
standard model, the MSSM. In this paper we address some aspects of
this model and of its extension to include neutrino masses. In
particular, we analyze the implications of neutrino masses to a new
mechanism recently introduced by Dimopoulos, Giudice and Tetradis
(DGT) \cite{ref:dgt} to ameliorate the flavor problem of the MSSM.
In this analysis we present their mechanism somewhat differently
from their original
work, and then focus on its implications to rare leptonic processes
such as $\mu\to e\gamma$. If the DGT mechanism is operative at a
very large momentum scale $\Lambda$, then data about such rare
processes can be combined with results from direct SUSY searches and
from various neutrino experiments to reveal important information
about the leptonic couplings at the scale $\Lambda$.

The long-standing problem which DGT have sought
to solve is that of flavor-changing neutral currents (FCNC) in the
MSSM. If the soft mass matrices of squarks and sleptons are---as
expected---not too far above the electroweak scale, and if they are
neither proportional to unit matrices nor aligned with the
corresponding fermion mass matrices, then they induce unacceptably
large contributions to various FCNC processes, in particular neutral
kaon oscillations and $\mu \to e \gamma$. Various mechanisms have
been suggested to overcome this difficulty: if the gauginos are
somewhat heavier than expected, they would raise the squark masses
and make them roughly proportional to the unit matrix (though the
difficulties in the leptonic sector would be harder to overcome); if
the soft
masses start out universal---proportional to the unit matrix---at a
very high scale, they typically remain so in their first- and
second-generation entries, and so contribute little to the two
sensitive FCNC processes mentioned above; and if some flavor
symmetries align the squarks with the quarks and the sleptons with
the leptons, then once again the FCNC contributions can be
suppressed. But the first two solutions have serious shortcomings:
gaugino dominance requires unnaturally heavy gauginos and moreover
is not very effective for $\mu \to e \gamma$; and universal soft
masses seem an unlikely outcome of various theories at the highest
scales.
The third approach postulates a set of approximate symmetries to
explain both the observed fermion mass matrices and their alignment
with the squark and slepton masses \cite{ref:hor}. It is somewhat
similar in spirit to the approach of DGT, in that the same
suppression mechanism which works in the quark and lepton sectors is
applied to the SUSY-breaking sector, and generically yields similar
FCNC suppression. The actual amount of suppression, though, varies
considerably depending on the horizontal symmetries used, and can be
stronger or weaker than in the DGT scenario.  In any case, a
thoroughly novel mechanism for
suppressing FCNC's is very welcome. The recent proposal of DGT
introduces just such a mechanism: a model, or more correctly a
paradigm, in which the squark and slepton mass matrices are
dynamically aligned with those of the corresponding fermions.

In this letter we first present the idea and the assumptions of
DGT somewhat differently than in the original
proposal, focusing on the intrinsic link between any such dynamics
and various fundamental concerns about the vacuum energy. We then
show that even if such a mechanism is viable, and can greatly
improve the situation in the quark sector, we do not expect it to be
nearly sufficient in the lepton sector. In the paper of DTG
individual lepton numbers were conserved, and therefore not
surprisingly the slepton and lepton mass matrices were exactly
aligned and no FCNC processes such as $\mu \to e \gamma$ could
occur. However, realistically we expect the lepton number symmetries
to be violated in the neutrino sector for a variety of
phenomenological reasons. As we show below, such violations will
induce a misalignment between leptons and sleptons. Though we can
not
at present predict the degree of misalignment precisely,
we expect the effect to be phenomenologically important, and to
yield valuable information about the leptonic flavor violations at
very high energies.

In the standard model, flavor violation comes about exclusively
through the Cabibbo-Kobayashi-Maskawa mixing matrix. We can choose a
basis for the quark fields such that the gauge interactions are
flavor-conserving, as are the Yukawa couplings of the leptons $Y_E =
\widehat Y_E$ (the hat indicates that the matrix is diagonal) and of
the down-type quarks $Y_D = \widehat Y_D$, but then there is no more
freedom to diagonalize the Yukawa couplings of the up-type quarks
$Y_U$:
they are given by $Y_U = K^\dagger \widehat Y_U$, where
$K$ is the CKM matrix. Since the Yukawa couplings of the first
two quark generations are small and the mixing with the third
generation is small, the standard model exhibits very feeble FCNCs.
In the lepton sector, flavor is exactly conserved.

The minimal extension of this standard model introduces eight more
potentially flavor-violating matrices: the five scalar mass matrices
$\tilde m_Q^2$, $\tilde m_U^2$, $\tilde m_D^2$, $\tilde m_L^2$, and
$\tilde m_E^2$, and the three trilinear scalar coupling matrices
$A_E$, $A_D$, and $A_U$. We will choose once again to keep the gauge
and gaugino interactions flavor-diagonal, and to do so we always
rotate superpartners together. If we then stay with the above choice
of basis for quark and lepton fields, we have no more freedom to
diagonalize the eight new soft-SUSY-breaking matrices. If their
off-diagonal terms are not suppressed relative to the diagonal ones,
unacceptably large FCNCs can arise, as discussed above. We will
concentrate first on the scalar masses, and then return to a
discussion of the $A$ terms.

DGT have proposed that these scalar mass matrices
be promoted to dynamical fields rather than be treated as mere
parameters. The advantage is that there may then exist a dynamical
relaxation mechanism which would align these matrices with the
Yukawa matrices and thereby minimize the flavor-changing
interactions.  Such a situation may arise in string theory, where
the low-energy field theory parameters are often dynamically
determined by the vacuum expectation values of certain fields. The
fundamental, microscopic theory and the low-energy effective field
theory are matched at a scale $\Lambda$ which we take to be of order
the string or Planck scales, but could
also be some lower scale. The Yukawa couplings are assumed to
be fixed by the fundamental theory, perhaps by expectation values of
fields with very large masses, so they are simply parameters of the
effective theory. As for the scalar mass matrices, it is conceivable
that
their eigenvalues and orientations are determined by different
mechanisms.  We will only consider the ``disoriented'' scenario in
which the eigenvalues are first fixed by some dynamics responsible
for supersymmetry breaking, and then the orientations are
dynamically determined by a set of light moduli fields. (We call
them ``moduli'', with an abuse of language, because they would
correspond to flat directions when either the Yukawa couplings
vanish or supersymmetry is unbroken.) If we
denote the scalar masses by $3\times 3$ matrices $\tilde m_I^2$
where $I$ runs over the five fields $Q$ (squark doublets), $U$
(up-type antisquark singlets), $D$ (down-type antisquark
singlets), $L$ (slepton doublets) and $E$ (charged slepton
singlets), and diagonalize them by means of unitary matrices $U_I$
as in Refs.~\cite{ref:dgt},
\begin{equation}
\tilde m_I^2 = U_I^{\dagger} \widehat m_I^2 U_I\,,\qquad I =
Q,U,D,L,E\,,
\label{eq:uadef}
\end{equation}
where $\widehat m_I^2$ is a real diagonal matrix, then the
disoriented assumption amounts to promoting $U_I$ to
dynamical fields, whose expectation value is determined
by minimizing an effective potential. One could further expand the
set of dynamical fields to include the eigenvalues $\widehat m_I^2$,
resulting in the ``plastication'' scenario of DGT, but we limit our
discussion of this scenario to a brief remark towards the end of
this work.

To determine what is the physics which fixes the alignment of the
$U_I$, we need to examine the exact effective potential of the
theory, which includes the effects of quantum fluctuations from all
scales. After allowing for possible new physics beyond the MSSM at
some intermediate SUSY-invariant scale $M$ between the cutoff
$\Lambda$ and the effective SUSY-breaking scale $\tilde m \sim m_Z$
in the observable sector, we may write the effective potential
generically as
\begin{equation}
V_{\rm eff} = c_4 {\cal O}(\Lambda^4) + c'_4 {\cal O}(\Lambda^2 M^2)
+ c''_4 {\cal O}(M^4) + c_2 {\cal O}(\Lambda^2 \tilde m^2) + c'_2
{\cal O}(M^2 \tilde m^2) + c_0 {\cal O}(\tilde m^4)  + \ldots
\label{eq:veff}
\end{equation}
where we have omitted a constant and any terms smaller than $\sim
\tilde m^4$. Since the first
three terms do not involve any supersymmetry breaking, they must
vanish inasmuch as the vacuum energy of a supersymmetric theory
vanishes: $c_4 = c'_4 = c''_4 = 0$. Assuming a hierarchy $\Lambda
\gg M \gg \tilde m$, the dominant term would then be the ${\cal
O}(\Lambda^2 \tilde m^2)$ part, which appears as a quadratic
divergence in the low-energy effective theory.  We will discuss such
divergences, and the possibility that they are absent, further
below, and argue
that not only are they expected, but that even if they vanish the
main results of our work will not change. Therefore, we will assume
that the ${\cal O}(\Lambda^2 \tilde m^2)$ terms dominate and
determine the orientation of the scalar masses. We note in passing
that whatever mechanism is ultimately responsible for cancelling the
cosmological constant, namely the constant term in $V_{\rm eff}$,
may have implications for the other terms in this potential, but we
cannot yet speculate on what those implications may be.

The direct consequence of this assumption, as noted already in
Ref.~\cite{ref:dgt}, is that the physics which determines
the alignment is the physics at the cutoff scale $\Lambda$, namely,
just at the scale in which the effective low-energy theory breaks
down. All higher-dimension ``irrelevant'' operators are in principle
as relevant as the renormalizable ``relevant'' ones. (For instance,
operators with four derivatives and a non-trivial flavor structure
contribute to $c_2$ already at 1-loop order, potentially competing
in an important way with the aligning ``force'' determined by the
low energy Yukawa couplings.) Therefore it
is really the entire fundamental theory at the scale $\Lambda$,
rather than just its low-energy sector (the MSSM plus any additional
new physics below $\Lambda$), which sets the dynamics of the scalar
mass orientations. Since our experimental knowledge is limited to
the low-energy sector while our theoretical understanding of the
fundamental theory is not sufficiently advanced to calculated
$V_{\rm eff}$, we cannot proceed further without some strong
assumptions. This is a serious and apparently inherent weakness of
this approach to solving the flavor problem. However, it can also be
viewed favorably as affording us a window into the fundamental
theory at the scale $\Lambda$: sensitivity to such scales means that
our predictions are not independent of this unknown realm, and
therefore that they may be used to experimentally probe it.

What, then, can we say about the ${\cal O}(\Lambda^2 \tilde m^2)$
terms? First, consider the radiative contribution of the
MSSM modes. The MSSM would possess a global $\rm U(3)^5$ flavor
symmetry if not only the dynamical fields $\tilde m_I^2$ would
transform
under this symmetry but also the Yukawa couplings would transform
appropriately. Since the Yukawa couplings are in fact fixed
parameters, they are the spurions which carry the information about
$\rm U(3)^5$
breaking. Hence, to lowest order in these Yukawa couplings, the MSSM
modes
contribute
\begin{eqnarray}
V_{\rm eff}^{\rm MSSM} &=& {\Lambda^2 \over \left(16\pi^2\right)^2}
\left[ c_Q {\rm Tr}\,\, \tilde m_Q^2 \left(
  K^{\dagger} \widehat Y_U \widehat Y_U^{\dagger} K +
  k_Q \widehat Y_D \widehat Y_D^{\dagger} \right) +
\right.\nonumber\\
& &   \phantom{\Lambda^2 \over \left(16\pi^2\right)^2}\,\,\,
       c_U {\rm Tr}\,\, \tilde m_U^2
  \widehat Y_U^{\dagger} \widehat Y_U +
       c_D {\rm Tr}\,\, \tilde m_D^2
  \widehat Y_D^{\dagger} \widehat Y_D + \label{eq:veffmssm}\\
& &    \phantom{[\Lambda^2 \over \left(16\pi^2\right)^2}\,\,\left.
       c_L {\rm Tr}\,\, \tilde m_L^2
  \widehat Y_E \widehat Y_E^{\dagger} +
       c_E {\rm Tr}\,\, \tilde m_E^2
  \widehat Y_E^{\dagger} \widehat Y_E\right]
\nonumber
\end{eqnarray}
to the full effective potential. The $c_I$ and also $k_Q$ are
numerical (scalar) coefficients which can only be calculated once
the matching
conditions are specified at the cutoff scale. Fortunately, we only
need to assume that they do not vanish. We also expect $k_Q$ to be
of order one; indeed in the low energy MSSM, $k_Q=1$ to lowest order
in the Yukawa couplings and up to small hypercharge effects. Then
these MSSM
contributions align $\tilde m_U^2$ with the {\it diagonal}
mass-squared matrix $\sim Y_U^{\dagger} \widehat Y_U$ of the up-type
quarks, $\tilde m_D^2$ with the {\it diagonal} mass-squared matrix
$\sim Y_D^{\dagger} \widehat Y_D$ of the down-type quarks, and
$\tilde m_Q^2$ with the {\it non-diagonal} linear combination
$K^{\dagger} \widehat Y_U \widehat Y_U^{\dagger} K + k_Q \widehat
Y_D \widehat Y_D^{\dagger}$. In the leptonic sector, there is only
one spurion, the diagonal Yukawa matrix $Y_E$, so both $\tilde
m_L^2$ and $\tilde m_E^2$ align with it and become diagonal:
individual lepton numbers are conserved.

Operators with higher powers of the Yukawa couplings will not
significantly change the minimum configuration of the sfermion
masses. This observation, based upon direct inspection of such
operators, is independent of any additional suppressions from powers
of $1/16\pi^2$. One underlying reason is that the minimum
configuration corresponds to points of enhanced flavor symmetry,
thus off diagonal entries in the configuration will always be
proportional to the appropriate CKM mixings violating those
symmetries. Additional suppressions arise because the Yukawa
coupling eigenvalues are hierarchical and mostly $\ll 1$. For
example, the $i,j$ entry of the matrix to which $\tilde m^2_U$
aligns will be of order $(m_u)_i (m_u)_j \theta_{ij}$ (in order to
account for the chiral quantum numbers of the $U$ multiplet); this
matrix is close to being diagonal because when $i < j$, $(m_u)_i
(m_u)_i \ll (m_u)_i (m_u)_j \ll (m_u)_j (m_u)_j$, and also
$\theta_{ij} \ll 1$.

Of course, the MSSM modes are but a small part of the theory at the
cutoff scale, and we have already seen that it is this full
fundamental theory which determined the alignment of the scalar
masses. To make progress, therefore, we must make some strong
assumption about the remaining physics. If it is completely
arbitrary, the partial alignment which would result from the
low-energy modes is generically destroyed. On the other hand, if the
remaining physics is related to the Yukawa couplings, in the sense
that it preserves the same approximate symmetries, then the
alignment can be preserved. This will therefore be our working
assumption:
\begin{itemize}
\item the {\em explicit} (rather than spontaneous) flavor violations
in the theory at the cutoff are {\em entirely} parametrized by the
Yukawa couplings, treated as spurions.
\end{itemize}
It follows from this {\it minimality assumption} that the
complete effective action has the same form as
Eq.~(\ref{eq:veffmssm}) but with different coefficients, $c_I \to
c'_I$ and $k_Q \to k'_Q$. The effective action will be minimized
when the soft masses are as closely aligned as possible with the
Yukawa couplings, that is, when the approximate flavor symmetries
are maximized. (The ground state of many physical systems is the
state of enhanced symmetry, so our results may have quite a wide
range of applicability.) The minimality assumption itself is very
restrictive, and hence quite predictive. Deviations from our
predictions would indicate that
there are new flavor violations in the cutoff theory which would
normally be almost inaccessible to experimental probes. We will make
this assumption throughout this work, and derive some quantitative
phenomenological consequences. It should be noted, however, that
even if there are
other sources of flavor violation which misalign the scalar masses,
they are unlikely to cancel the ones we can calculate from our
effective potential, and hence our predictions serve very generally
as rough lower
bounds on the expected experimental signals.

Under the minimality assumption, the mass matrix of the squark
doublets aligns with the linear combination $K^{\dagger} \widehat
Y_U
\widehat Y_U^{\dagger} K + k'_Q \widehat Y_D \widehat
Y_D^{\dagger}$.
If we assume $k'_Q \sim {\cal O}(1)$ as suggested by $V_{\rm
eff}^{\rm MSSM}$, then the second term in the linear combination may
be ignored relative to the first, and $m_Q^2$ aligns approximately
with the up-type Yukawa couplings, that is, it is misaligned by the
CKM matrix $K$ relative to the diagonal down-type quark masses. Thus
the strength of FCNCs in the quark-squark sector is suppressed by
the
small off-diagonal matrix elements of $K$, and this is the
extent to which the disorientation mechanism alleviates the flavor
problem in this sector. (Recall that, unlike the orientations, the
eigenvalues of the squark mass matrices are fixed parameters in our
current non-plasticated discussion.) Disorientation sets the generic
values of such quantities as $\epsilon_k$, $\epsilon'/\epsilon_K$
and $B\bar B$ and $D \bar D$ mixings at (or below) their
experimental values or bounds. On the other hand
for the $K_L- K_S$ mass difference it is less
effective\cite{ref:dgt}, requiring a further suppression
\begin{equation}
{\left(\tilde m_{Q2}^2 - \tilde m_{Q1}^2\right) \over \tilde m_Q^2}
< 0.1 \left({\tilde m_Q \over 300\,\rm GeV}\right)\,.
\label{eq:klks}
\end{equation}
We see that the disorientation mechanism suffices for satisfying
most phenomenological bounds, but that
some other solution---such as an accidental approximate degeneracy,
an
approximate universality, or perhaps plastication
\cite{ref:dgt}---is still needed, at least for $\Delta m_K$. (The
usual bounds on supersymmetric parameters from $K_L - K_S$ mass
difference are not greatly alleviated by disorientation simply
because disorientation can only suppress flavor-changing neutral
currents by a factor of the appropriate CKM angle, but a much
greater suppression is mandated by the experimental value of $\Delta
m_K$, and is provided in the Standard Model by the lightness of the
charm quark. Other quantities, such as $\epsilon_K$, which are
sufficiently suppressed in the standard model by small CKM mixing
angles, are indeed similarly suppressed in a disoriented scenario.)

What of the leptonic sector? It is well-known that for generic soft
masses the bounds on FCNCs from the radiative $\mu$ decay process
$\mu \to e \gamma$ are
considerably stronger than the $K_L - K_S$ bounds. Under the
minimality assumption, the alignment is essentially
determined by the Yukawa couplings of the low-energy modes, which in
the MSSM conserve individual lepton numbers. Therefore
\cite{ref:dgt} all FCNCs in the lepton sector vanish.
However, there is considerable evidence that individual lepton
numbers are not in fact conserved in nature, and therefore that
there is indeed low-energy physics beyond the MSSM. The main
indication comes from  the solar neutrino flux, whose observed
deficit can be explained by resonant (MSW) neutrino oscillations
\cite{ref:msw}
favoring neutrino masses in the $10^{-3}$ eV range
\cite{ref:neutmass}.
Other, perhaps less compelling, indications come from the
atmospheric neutrino problem and from the density fluctuations at
large scales measured by COBE \cite{ref:cobe}. This latter
observation can be more easily explained
provided a substantial fraction of the dark matter is hot
\cite{ref:fluct,ref:hot}. With the MSSM particle content the only
way to obtain that is to have some neutrino mass (presumably the
$\nu_\tau$) in the eV range. Once we allow for flavor violation in
the leptonic sector, then it is bound
to show up in the slepton mass matrices. Almost all models for
neutrino masses involve a
Majorana mass matrix $M_N$ for the singlet (``right-handed'')
neutrino states, which also interact with the lepton doublets
through a new set of Yukawa couplings $Y_N = K_L^\dagger \widehat
Y_N$; here we use the basis defined above in which $Y_E = \widehat
Y_E$ is diagonal, so $K_L$ is the leptonic analog of the CKM matrix
$K$. In the quark sector there is a natural choice for which up-type
quark to group with a given down-type quark in a single generation:
one defines the three generations so as to make the CKM matrix close
to the identity. In the leptonic sector we will find it convenient
to define $K_L$, in the above basis, as the matrix which brings
$Y_N$ into a diagonal form with {\it increasing} diagonal entries.
As we will see below, the slepton masses will be aligned with $K_L$,
so to avoid excessive flavor-changing processes $K_L$ will have to
be close to the identity up to a permutation matrix. For simplicity
we will assume that this permutation matrix is the identity. The
eigenvalues of
$M_N$ are much larger than the weak scale, making the
observed neutrinos mostly left-handed and very light via the see-saw
mechanism. Leptonic flavor violation is parametrized by the
misalignment matrix $K_L$ and by $M_N$. We will assume, in the
spirit of our previous minimality assumption and in agreement with
phenomenological expectations, that $M_N \ll \Lambda$ (where again
$\Lambda$ is of order the Planck or string scale). Therefore the
effective potential for the slepton masses will be of the form (to
lowest order in the Yukawa couplings)
\begin{eqnarray}
V_{\rm eff}^{\rm leptonic} &=& {\Lambda^2 \over
\left(16\pi^2\right)^2}
\left[ c'_L {\rm Tr}\,\, \tilde m_L^2 \left(
  K_L^{\dagger} \widehat Y_N \widehat Y_N^{\dagger} K_L +
  k'_L \widehat Y_E \widehat Y_E^{\dagger} \right) +
\right.\nonumber\\
& &   \phantom{\Lambda^2 \over \left(16\pi^2\right)^2}\,\,\left.
       c'_N {\rm Tr}\,\, \tilde m_N^2
  \widehat Y_N^{\dagger} \widehat Y_N +
       c'_E {\rm Tr}\,\, \tilde m_E^2
  \widehat Y_E^{\dagger} \widehat Y_E \right]\,.
\label{eq:vefflept}
\end{eqnarray}
The SU(2)-singlet charged slepton masses align to lowest order with
the mass-squared matrix of the charged leptons, and hence are
diagonal in the basis we have chosen:
\begin{equation}
\tilde m_E^2 \simeq \widehat m_E^2\,.
\label{eq:mealign}
\end{equation}
But the presence of the right-handed neutrinos at the cutoff scale,
and
therefore of the Yukawa couplings $Y_N$ as spurions violating
leptonic flavor symmetries, is enough to misalign the SU(2)-doublet
slepton mass matrices. The misalignment is frozen in at the scale
$\Lambda$, and so it remains even after the right-handed modes are
integrated out of the low-energy theory. The sensitivity to the
theory at the cutoff scale implies an insensitivity to the details
of the decoupling of the right-handed neutrinos and to the flavor
violation in $M_N$. Moreover, as in the quark sector, we will assume
that $k'_L \sim {\cal O}(1)$ and that the Yukawa couplings of the
neutrinos are larger than those of the charged leptons:  then
$m_L^2$ aligns to a good precision with the Yukawa couplings of the
right-handed neutrinos and therefore is misaligned by the leptonic
CKM matrix $K_L$ relative to the charged-lepton masses,
\begin{equation}
\tilde m_L^2 \simeq K_L^{\dagger} \widehat m_L^2 K_L\,,
\label{eq:mlalign}
\end{equation}
much as $\tilde m_Q^2$ was misaligned by the CKM matrix $K$ relative
to the down-type quark masses. Consequently, the strength of
leptonic FCNCs is sensitive only to the matrix $K_L$ and not to the
overall unknown size of the neutrino Yukawa couplings. By measuring
processes such as $\mu \to e \gamma$ (and $\mu\to 3e$) and $\tau \to
\mu \gamma$, we can directly probe these leptonic mixing angles. We
stress that the
above result does not necessarily require Yukawa couplings $\sim 1$
in the neutrino sector---having $Y_N \roughly > Y_E$ suffices.

Another potential source for flavor violations is the set of
$A$-term matrices. As discussed in Ref.~\cite{ref:dgt}, various
assumptions could be made about the $A$ terms: they are at once
similar to the Yukawa couplings, since they couple ``left''- and
``right''-handed modes, and to the scalar masses, since they break
supersymmetry softly. If the orientation as well as the eigenvalues
of the $A$ terms were fixed at the cut-off scale (like the Yukawa
couplings), then to satisfy phenomenological bounds they would need
to either be negligibly small or closely aligned with the Yukawa
couplings; this would also follow from our minimality hypothesis
(which would require $A_i \propto Y_i$ or $A_i = 0$), and is also
one component of the commonly-made universality assumption. (Notice
that fixed $A$-term orientations allow the $U_I$ soft-masses
orientations to be regarded as pseudo-Goldstone bosons of the flavor
symmetries.) Alternatively, the $A$ terms may have fixed eigenvalues
but dynamically-determined orientations (like the soft masses):
\begin{equation}
A_i = V_i^{\dagger} \widehat A_i W_i\,,\qquad i = U,D,N,E.
\label{eq:widef}
\end{equation}
Without loss of generality, we may arrange the (diagonal) entries of
$\hat A_i$ in ascending order. We will assume in particular that the
$V_i$ and $W_i$ fields are independent of the $U_I$ fields. (If they
were all regarded strictly as pseudo-Goldstone bosons of flavor
symmetries then there would be fewer independent fields---and
consequently insufficient freedom to align both the sfermion masses
and the $A$ terms.) What of the fixed eigenvalues $\widehat A_i$? If
they are too large, undesirable minima develop \cite{ref:dersav} in
the full MSSM scalar potential which spontaneously break the
electromagnetic gauge symmetry. The limiting values for $\widehat
A_i$ (unless the sfermions are extremely heavy) are roughly the
corresponding {\it fermion} masses:
\begin{equation}
\widehat A_i \,\roughly{<}\, m_i\,, \qquad i = u,d,n,e.
\label{eq:aibound}
\end{equation}
Thus we must suppose---as is always done---that the $A$ terms have
sufficiently small eigenvalues, at most comparable to the
hierarchical masses of the charged leptons. If in fact the
eigenvalues are comparable to the hierarchical charged lepton
masses, then they must also be at least roughly aligned, rather than
antialigned, with the corresponding leptons in order to satisfy
Eq.~(\ref{eq:aibound}). We now turn to the issue of alignment,
namely the expectation values of $V_i$ and $W_i$.

The alignment of the $A_i$ is determined under the minimality
assumption by a spurionic analysis similar to that for the sfermion
masses. To proceed, we further postulate that the only parameters
breaking the $\rm U(1)_R$ symmetry of the MSSM are the soft ones,
namely $A_i$, the bilinear Higgs coupling $B$, and the gaugino mass
parametrized by $M_{1/2}$, all of which transform in the same way
under R.  This symmetry will allow only terms involving $AA^\dagger$
and $A^\dagger M_{1/2}$ but not any $AA$ terms. Then the operators
determining
the alignment of $A_E$ are given to lowest order by
\begin{equation}
V_{\rm eff}^{\rm A} \sim
c^{\prime A}_{N} M_{1/2}^\dagger {\rm Tr}\,A_N Y_N^\dagger +
c^{\prime A}_{E} M_{1/2}^\dagger {\rm Tr}\, A_E Y_E^\dagger +
c^{\prime A}_{EN} {\rm Tr}\, A_E Y_E^\dagger Y_N A_N^\dagger  +
c^{\prime A}_{EE} {\rm Tr}\, A_E A_E^\dagger Y_N Y_N^\dagger + {\rm
h.c.}
\label{eq:veffa}
\end{equation}
What alignment do these terms induce?  The first operator, ${\rm
Tr}\,A_N Y_N^\dagger$, will likely dominate the alignment of $A_N$.
It can be positive or negative, and has its largest magnitude when
$W_N \simeq \unity$ (the unit matrix) and $V_N \simeq K_L$; hence
for $c^{\prime A}_{N}$ of any sign, the first term in the potential
is always minimized when $A_N \simeq K_L^\dagger \widehat A_N$ (up
to an overall sign). Similarly, the second operator is minimized
when $V_E \simeq W_E \simeq \unity$ so it favors $A_E \simeq
\widehat A_E$. Next, to analyze the third term, we recall that $V_N
\simeq K_L$ and $W_N \simeq \unity$, that $\widehat A_N$ and
$\widehat Y_N$ are arranged in increasing order, and that $\widehat
A_E$ and $\widehat Y_E$ are strongly hierarchical (unless the former
is negligibly small); then, up to terms of order $K_{Lij} m_i/m_j$
for $i < j$, the third term is minimized when $A_E \simeq \widehat
K_L^\dagger A_E$. Finally, the last operator in this potential is
positive semidefinite, and attains its largest magnitude when $V_E
\simeq K_L$. So if the coefficient $c^{\prime A}_{EE}$ is
nonnegligible, it must be negative in order to lead to approximate
alignment (rather than antialignment) of the charged sleptons with
the charged leptons.

Therefore the leptonic $A$ term relevant at low energies has the
form
\begin{equation}
A_E \simeq \tilde V_E^\dagger \widehat A_E
\label{eq:aealign}
\end{equation}
where the fixed matrix $\tilde V_E$ is, up to phases, $\simeq
\unity$ if the second operator in $V_{\rm eff}^{\rm A}$ dominates
over the second and
third, and otherwise has entries comparable to $K_L$. We comment
below on the possible CP-violating effects of these phases.

We should add that the above form for $A_E$ is valid at the cut-off
scale $\Lambda$. RG evolution to low energies will add two types of
terms to the cut-off expressions: one from the gauge sector and the
other from the Yukawa sector. The gauge contribution to $A_E$ is a
diagonal matrix proportional to $M_{1/2} Y_E$ (we will use an
approximate proportionality constant of $-0.3$). The gauge
contribution to the soft masses adds universal terms $\propto
M_{1/2}^2 \unity$, but these will not change the form of the mass
matrices. Note that, in accordance with low-energy measurements of
the gauge couplings and with the MSSM RG equations, we have assumed
that the three gauge couplings approximately unify near the cut-off
scale and that the gauginos have a common mass $M_{1/2}$ at that
scale. The Yukawa sector contributions were first partially analyzed
in Ref.~\cite{ref:bm} and were recently analyzed in detail in
Ref.~\cite{ref:hmty}. We will comment on these contributions briefly
below.

Before discussing the phenomenology let us recollect our
assumptions. First, we assumed that the slepton mass matrix
orientations were dynamical degrees of freedom fixed by a cut-off
scale effective potential in which the flavor violations are
entirely due to the Yukawa couplings (the minimality assumption),
and second, when right handed neutrinos are added to the MSSM
to account for neutrino masses, we assume that $Y_N$
are at least as large (in a matrix sense) as $Y_E$. The result is
Eqs.~(\ref{eq:mealign}) and (\ref{eq:mlalign}) for slepton masses:
namely, the ``left-handed'' [that is, SU(2)-doublet] soft slepton
masses $\tilde m_L^2$ are misaligned relative to the charged lepton
masses by the leptonic CKM matrix $K_L$, while the ``right-handed''
soft slepton masses $\tilde m_E^2$ are closely aligned with the
charged leptons. If the $A$ terms are not negligible, then we
further assumed that their orientations are dynamical according to
Eq.~(\ref{eq:widef}), that their eigenvalues satisfy
Eq.~(\ref{eq:aibound}), that alignment rather than antialignment
results from the effective potential, and that $K_L$ is close to the
identity. The last three requirements simply allow the stability of
the electroweak vacuum.  The resulting $A$ terms are misaligned on
their left-hand side by at most $\sim K_L$ and are aligned with the
charged leptons on their right-hand side.

With the low-energy soft SUSY-breaking parameters at hand, we may
study the expected phenomenology and compare the results to current
experimental bounds and to future experimental potential. We will
concentrate on the radiative flavor-changing muon decay $\mu \to e
\gamma$ since it furnishes perhaps the most sensitive probe of these
parameters, and since we anticipate its sensitivity to be greatly
improved in the near future. With the above form for the soft terms,
the radiative decay is completely dominated by one helicity
amplitude ${1\over2} {\cal A}_{R \to L} \bar e_L \sigma_{\mu\nu}
F^{\mu\nu} \mu_R$, while the other helicity
amplitude vanishes to lowest order in $m_e/m_\mu$. We have
independently and fully computed the dominant helicity amplitude in
the MSSM, neglecting terms of order $m_e/m_\mu$, and compared our
results with previous calculations\footnote{Much of the existing
literature omits parts of the amplitude, and in particular the
important contribution ${\cal A}_C$ arising from chargino
propagation with a mass insertion on the internal higgsino-wino
line. In comparing our results with two of the complete
calculations, we found one minor sign and normalization discrepancy
with Ref.~\cite{ref:sutter} and agreement with
Ref.~\cite{ref:hmty}.} We assume for brevity that mixing between the
first two generations dominates; the generalization to full
three-generation mixing is straightforward. Our result for the
branching ratio is $\br(\mu \to e \gamma) = {\displaystyle{\tau_\mu
m_\mu^3\over 16 \pi}} \left|{\cal A}_{R \to L}\right|^2$, where
${\cal A}_{R \to L} = {\displaystyle{e\over 16\pi^2}} m_\mu
K_L^{1\mu*} K_L^{1e}  \left[{\cal A}_A + {\cal A}_B +{\cal A}_C
+{\cal A}_D +{\cal A}_E\right]$, and
\begin{eqnarray}
{\cal A}_A &=& \half \sum_{i=1}^4
\left(g^{\prime 2} \left|U_{i1}\right|^2 +
      g_2^2 \left|U_{i2}\right|^2 +
      2 g' g_2 \,{\rm Re}\, U_{i1} U_{i2}^*\right)
\left[{f(M_{0i}^2/\tilde m_{L1}^2) \over \tilde m_{L1}^2} -
      (L1 \leftrightarrow L2) \right]
\label{eq:ampa} \\
{\cal A}_B &=& -g_2^2 \left(c_+^2 \left[
{g(M_{+1}^2/\tilde m_{L1}^2) \over \tilde m_{L1}^2} - (L1
\leftrightarrow L2)
\right] + s_+^2 \left[
{g(M_{+2}^2/\tilde m_{L1}^2) \over \tilde m_{L1}^2} - (L1
\leftrightarrow L2)
\right]\right) \phantom{A}
\label{ampb} \\
{\cal A}_C &=& {g_2^2\over v_D} \left(C_1 \left[
{j(M_{+1}^2/\tilde m_{L1}^2) \over \tilde m_{L1}^2} - (L1
\leftrightarrow L2)
\right] + C_2 \left[
{j(M_{+2}^2/\tilde m_{L1}^2) \over \tilde m_{L1}^2} - (L1
\leftrightarrow L2)
\right]\right) \phantom{A}
\label{ampc} \\
{\cal A}_D &=& -{g_2\over \sqrt{2}v_D} \sum_{i=1}^4 U_{i3}^* U_{i2}
M_{0i}
\left[{h(M_{0i}^2/\tilde m_{L1}^2) \over \tilde m_{L1}^2} -
      (L1 \leftrightarrow L2) \right]
\label{ampd} \\
{\cal A}_E &=& -\sum_{i=1}^4 \left(
   U_{i1}^2 g^{\prime 2} + U_{i1} U_{i2} g' g_2\right)^* M_{0i}
\times \label{ampe}\\
& & \qquad \left\{K_- \left[
{1\over \tilde m_{L1}^2-m_{E2}^2}
\left({h(M_{0i}^2/\tilde m_{L1}^2)\over\tilde m_{L1}^2} -
      {h(M_{0i}^2/\tilde m_{E2}^2)\over\tilde m_{E2}^2} \right)
 - (L1 \leftrightarrow L2)
\right] +\right. \nonumber\\
& & \qquad \left.\phantom{\{}K_+ \left[
{1\over \tilde m_{L1}^2-m_{E2}^2}
\left({h(M_{0i}^2/\tilde m_{L1}^2)\over\tilde m_{L1}^2} -
      {h(M_{0i}^2/\tilde m_{E2}^2)\over\tilde m_{E2}^2} \right)
 + (L1 \leftrightarrow L2)
\right] \right\} \nonumber\\
\end{eqnarray}
The overall factor $K_L^{1\mu*} K_L^{1e}$ is the off-diagonal mixing
in the leptonic CKM matrix; $U$ is the matrix which diagonalizes the
neutralino mass matrix $M_0$ via
\begin{eqnarray}
U M_0 U^\dagger &=& U
\left(\begin{array}{cccc}
M_1 & 0 & -g_1 v_D/\sqrt{2} & g_1 v_U/\sqrt{2} \\
0 & M_2 & g_2 v_D/\sqrt{2} & -g_2 v_U/\sqrt{2} \\
-g_1 v_D/\sqrt{2} & g_2 v_D/\sqrt{2} & 0 & \mu \\
g_1 v_U/\sqrt{2} & -g_2 v_U/\sqrt{2} & \mu & 0
\end{array}\right)
U^\dagger \label{eq:neutmass}\\
&=& \phantom{U}
\left(\begin{array}{cccc}
M_{01} & 0 & 0 & 0 \\
0 & M_{02} & 0 & 0 \\
0 & 0 & M_{03} & 0 \\
0 & 0 & 0 & M_{04}
\end{array}\right)\,;
\end{eqnarray}
$\tilde m_{Li}$ and $\tilde m_{Ei}$ are the mass eigenvalues of the
left-handed sleptons; $v_U$ and $v_D$ are the up- and down-type
Higgs boson mass parameters, satisfying $v_U^2 + v_D^2 = v^2 =
(174\,\rm GeV)^2$ and $v_U/v_D = \tan\beta$ (so the standard model
fermions have Dirac masses $m_i = Y_i v_{U,D}$); the chargino mass
matrix is diagonalized via
\begin{equation}
\left(\begin{array}{cc}
M_2 & g_2 v_D \\ g_2 v_U & -\mu
\end{array}\right) =
\left(\begin{array}{cc}
c_+ & -s_+ \\ s_+ & c_+
\end{array}\right)
\left(\begin{array}{cc}
M_{+1} & 0 \\ 0 & M_{+2}
\end{array}\right)
\left(\begin{array}{cc}
c_- & s_- \\ -s_- & c_-
\end{array}\right)
\end{equation}
from which we obtain the useful parameter combinations $C_1 = M_{+1}
s_- c_+/g_2$ and $C_2 = -M_{+2} s_+ c_-/g_2$; we also use $K_- = \mu
\tan\beta - 0.3 M_{1/2} + (\tilde A_{E2,1} + \tilde A_{E2,2})/2$ and
$K_+ = (\tilde A_{E2,1} - \tilde A_{E2,2})/2 \equiv \tilde A_{12}$
in which we expect $\tilde A_{E2,j} \equiv  (\hat A_{E2}/Y_\mu) (K_L
\tilde V_E^\dagger)^{j\mu*}/K_L^{j\mu*}$ to be between zero and the
SUSY-breaking scale, as discussed above; and the four loop functions
are defined via $f(x) = (2 x^3 + 3 x^2 - 6 x + 1 - 6 x^2 \ln x)/[12
(1-x)^4]$, $g(x) = (x^3 -6 x^2 +3 x + 2 +6 x \ln x)/[12 (1-x)^4]$,
$h(x) = (- x^2 + 1 +2 x \ln x)/[2 (1-x)^3]$, and $j(x) = (x^2 -4 x+
3 +2 \ln x)/[2 (1-x)^3]$.

Various contour plots of the calculated branching ratio for $\mu \to
e\gamma$  are shown in Fig.~1. In all the plots we have used a
slepton degeneracy $\Delta
\tilde m^2_L \equiv \tilde m_{L2}^2 - \tilde m_{L1}^2 = 0.1 \tilde
m_L^2$ and a leptonic mixing $\left|K_L^{1\mu*} K_L^{1e}\right| =
0.04$. The horizontal axis spans values of the $\mu$ parameter
between $-500\,\GeV$ and $500\,\GeV$, while the vertical axis spans
the same range of the approximately-unified gaugino mass $M_{1/2}$
at the cut-off scale. For various values of $\tan\beta$ and $\tilde
m_L$, the figures show contours of constant branching ratio
normalized to the current experimental upper bound of $\br_{\rm
exp}=4.9\times 10^{-11}$: the black, dark gray, light gray, and
white regions indicate $\br/\br_{\rm exp} < 0.1$, $0.1 <
\br/\br_{\rm
exp} < 1$, $1< \br/\br_{\rm exp} < 10$, and $10 < \br/\br_{\rm
exp}$,
respectively. Also shown, as cross-hatched regions, are those
parameter ranges excluded by LEP bounds on the lightest chargino and
neutralino masses. In the top row of plots, in which $\tan\beta = 2$
while $\tilde m_L$ varies between 100 GeV and 500 GeV, we have used
$\tilde A_{E2,1} = 0$ (which would result from $\tilde V_E = K_L$)
and $\tilde A_{E2,2} = 100\,\GeV$.  Making $\tilde A_{E2,1} \sim
\tilde A_{E2,2}$ would not change the results significantly. For the
remaining two rows of the figure, in which $\tan\beta = 2$ or
$\tan\beta = 5$, we have set $\tilde A_{E2,2} = 0$. Making $\tilde
A_{E2,2} = 100\,\GeV$ does not make much difference when $\tan\beta
= 5$ so we omit the corresponding figure. To properly interpret
these contours for any mixing, their scaling behavior is needed:
\begin{equation}
{\rm Br}(\mu \to e \gamma)= 4.9\times 10^{-11} \left [{{[\Delta
\tilde m^2_L+f M_0 \tilde A_{12}]/\tilde m_L^2}\over 0.1}\right
]^2\left [{|K_L^{1\mu*} K_L^{1e}|\over
0.04}\right ]^2 \left [{300\,\GeV\over \tilde m_L}\right ]^4  F
\label{eq:brratio}
\end{equation}
where $F$ and $f$ arise from loop functions.  Under different
assumptions about the mixing angles and degree of degeneracy, the
allowed regions will correspond to different contours in our plots.
The value of $F$ is $\sim 1$ when $\mu \sim M_{1/2} \sim \tilde
m_L$, but can be one or two orders of magnitude larger when $\mu$ or
$M_{1/2}$ are hierarchically lower than the slepton mass. Thus,
while light sleptons result as expected in very large branching
ratios (unless $\mu$ is accurately tuned to produce a cancellation),
simply raising the slepton masses without raising $\mu$ and
$M_{1/2}$ does not immediately lower the branching ratio: to quickly
suppress the branching ratio, the entire SUSY-breaking scale must be
raised, which necessitates fine-tuning the electroweak scale. The
amplitude responsible for this behavior is the oft-neglected ${\cal
A}_C$, which is never negligible and which depends logarithmically
on this hierarchy, dominating the  amplitude by a factor of $\sim
10$ when the sfermions are $\sim 3$ times as heavy as the charginos.
There is also a significant enhancement in the branching ratio when
$\tan\beta$ is large, as in many other processes which then require
suppression to agree with experiment \cite{ref:us}. Finally, the $A$
term contribution are often significant, especially when the
``left-handed'' soft-breaking masses are very nearly degenerate
$(\Delta \tilde m^2_L \ll \tilde m_L^2)$

The size of leptonic mixings we have inserted is
consistent with that suggested by the MSW solution of the solar
neutrino problem. Actually, the $K_L$ mixings and the neutrino
mixings observed at low energies (via neutrino oscillations of
various sorts) are in general only indirectly related, via the
Majorana mass matrix $M_N$. However, they are essentially equal when
the eigenvalues of $M_N$ are all of the same order while those of
$Y_N$ are strongly hierarchical.
With such mixings, Fig.~1 and
Eq.~(\ref{eq:brratio}) indicate that we need significant degeneracy
in the slepton masses and a somewhat high SUSY-breaking scale to
suppress $\mu\to e\gamma$ below its experimental bound---in fact,
roughly the same degeneracy and SUSY-breaking scale as were needed
to satisfy the neutral kaon mixing constraints. Thus the lepton
sector fares no better (and no worse) than the quark sector in a
disoriented scenario when neutrinos have sizeable Yukawa couplings
at the cut-off scale. Admittedly, we have the freedom in the
leptonic sector to assume that {\it for some unknown reason} the
charged and neutral leptons are very closely aligned at the cut-off
scale, in which case the mixing needed for the MSW scenario must be
provided by the Majorana mass matrix of the right-handed neutrinos.
Or we could assume that the Yukawa couplings of the neutrinos are
much smaller than those of the charged leptons, though this seems
unlikely. Otherwise, the SUSY-breaking scale must be at least
several hundred GeV, implying the usual fine-tuning problems for the
Z boson mass, or the left-handed slepton mass eigenvalues must be
made highly degenerate. If the leptonic mixings $|K_L^{1\mu*}
K_L^{1e}|$ are larger, say $\sim \sqrt{m_e/m_\mu} \simeq 0.07$ or
even $\sim \theta_c \simeq 0.2$, then the SUSY-breaking scale must
be raised further or the sleptons be made more degenerate to
accommodate the experimental bounds.

These arguments will be greatly strengthened by the planned
improvement in the experimental searches for $\mu\to e\gamma$.
Current proposals call for sensitivity to branching ratios as low as
$10^{-14}$ \cite{ref:cooper}. If the disoriented scenario is
correct, we certainly expect that $\mu\to e\gamma$ will be observed
in this next
generation of experiments. By measuring that rate we would gain some
direct information about the leptonic CKM matrix. Of course, the
other
parameters affecting the branching ratio must be measured as well,
but they will probably be determined within the coming decade.

The above predictions of the disoriented scenario should be
contrasted with the effects of RG evolution in the universal
scenario \cite{ref:bm,ref:hmty}. In the latter
flavor violating slepton masses vanish by fiat at the cut-off
scale, but are induced at lower scales by the neutrino Yukawa
couplings via RG evolution: the diagonal slepton masses $\tilde
m_0^2$ are augmented by $\delta \tilde m_L^2=(3/8\pi^2) \tilde m_0^2
Y_N Y_N^\dagger \ln(\Lambda/M_N)$, leading to a branching ratio
given by Eq.~(\ref{eq:brratio}) but with a mass splitting $\Delta
\tilde m_L^2/\tilde m_L^2=(3/8\pi^2) Y_{\nu_\mu}^2
\ln(\Lambda/M_N)$. Thus, unless the neutrino Yukawa coupling is
of order one, the slepton masses are highly degenerate and hence
this rare $\mu$ decay is greatly suppressed. The disoriented
scenario in effect allows $\Delta \tilde m_L^2$ to be a free
observable parameter while keeping the degree of misalignment
between leptons and sleptons small, namely $\simeq K_L$.

What does the disoriented scenario predict when $\tilde m_{L1}^2 =
\tilde m_{L2}^2$ for some reason, such as plastication
\cite{ref:dgt}? The only contributions to $\mu\to e\gamma$ are those
proportional to off-diagonal $A$ terms [namely the $K_+ =
(\tilde A_{E2,1} - \tilde A_{E2,2})/2$ term in ${\cal A}_E$] and
those
involving the third family left-handed sleptons. When $\tilde
A_{E2,1} - \tilde A_{E2,2} \simeq 100\,\GeV$ and the slepton mass is
$\simeq 250 \,\GeV$ the resulting rate is just below the present
bound; for fixed gaugino mass, the rate decreases with the eighth
power of the slepton mass, so heavy sleptons would only allow
detection at the next generation of experiments. To account for
mixing with the third family, the relevant mixing angle $K_L^{3\mu*}
K_L^{3e}$ should be substituted for $K_L^{1\mu*} K_L^{1e}$ in the
above calculation. Assuming these have the same size as their quark
sector counterpart, this contribution alone yields a branching ratio
roughly an order of magnitude below the current bound even if the
slepton masses are not degenerate and are $\sim 100\,\GeV$.

Our discussion so far has assumed that the orientations of squarks
and sleptons are independent dynamical degrees of freedom. In the
context of a grand-unified theory, the larger symmetry would
typically reduce the number of independent orientations.  As
discussed in Ref.~\cite{ref:dgt}, the result in a unified
disoriented model is a sleptonic mixing angle of order the Cabibbo
angle $\theta_c \simeq K^{1s*} K^{1d}$, or perhaps of order
$\sqrt{m_e/m_\mu}$ in a more detailed and realistic model. Since the
mixing angle is larger than the $0.04$ we used above, the branching
ratio of $\mu\to e\gamma$ is also larger. As a consequence we expect
that in a disoriented GUT scenario the superpartners are quite heavy
or the charged sleptons of the first two generations are highly
degenerate.  Third-generation and $A$ term effects are then
important, and may dominate if $\tilde m_{L1}^2 - \tilde m_{L2}^2$
is sufficiently small. In very simple disoriented GUT models the
misalignment may be only in the ``right-handed''  sector, but
generically it is present in both sectors. In fact, a disoriented
GUT scenario and a conventional GUT (see Ref.~\cite{ref:bh}
for a detailed analysis) have similar predictions.  They both differ
qualitatively, however, from the disoriented non-unified scenario in
an important way: the helicity of the amplitudes.  As we have shown,
in a non-unified disoriented scenario with only the MSSM fields plus
right-handed neutrinos at high scales, the process is completely
dominated by the single helicity amplitude $\mu_R\to e_L \gamma$. On
the other hand, in a realistic unified theory of flavor we expect
flavor violations of comparable order in both the left- and
right-handed sectors, while the minimal (and unrealistic) SU(5)
model produces only right-handed
mixing. Therefore in any unified theory we expect an amplitude for
$\mu_L\to e_R \gamma$ at least as large as $\mu_R\to e_L \gamma$.
Fortunately, in the planned experiments the decaying muon is
polarized, so if sufficiently many $\mu\to e\gamma$ are observed,
the angular distribution of the emitted electrons would reveal the
helicity of the amplitude. A pure $\mu_R\to e_L \gamma$ result would
be difficult to understand in a generic unified theory (disoriented
or otherwise), but would be expected in a disoriented scenario if
$\rm SU(3)\times SU(2)\times U(1)$ is the gauge group up to the
cut-off scale.

Another related rare $\mu$ decay is $\mu \to 3e$.
It was recently observed \cite{ref:hmty} that,
in contrast to previous statements in the
literature, the amplitude which dominates
this branching ratio is not the box diagram
but rather the (photon) penguin diagram.
Indeed, while the box contribution would
be several orders of magnitude below the
experimental bound, the penguin diagram
yields a branching ratio \cite{ref:hmty}
\begin{equation}
{\br(\mu\to 3e)\over \br(\mu\to e\gamma)} \simeq
{\alpha\over8\pi} \left({16\over3}\ln {m_\mu\over2m_e}
 - {14\over9}\right)
\simeq 0.36 {\br_{\rm exp.\,bound}^{\rm present}(\mu\to 3e)
\over  \br_{\rm exp.\,bound}^{\rm present}(\mu\to e\gamma)}
\label{eq:muee}
\end{equation}
which is comparable to the experimental bound
when $\mu\to e\gamma$ is close to its experimental
bound. At present, $\mu\to 3e$ yields slightly
weaker constraints than $\mu\to e\gamma$; only
if the precision of $\mu\to 3e$ experiments keeps
pace with the planned improvements in $\mu\to e\gamma$
searches will the former process remain competitive.
As pointed out in Ref.~\cite{ref:hmty}, the penguin
diagram is enhanced by $\ln(m_\mu/2m_e)$, which results
from phase space integration as an electron and positron become
collinear. (The coefficient of the log is just determined
by the  QED $\beta$ function by requiring the cancellation of
the infrared divergences in the inclusive rate to order $\alpha$.)
We should remark, however, that the experimental
resolution may not allow highly collinear $e^+e^-$ pairs
to be distinguished from other processes (including
$\mu\to e\gamma$!), so the the denominator in the log
should be replaced by the appropriate minimum resolvable energy.

We have assumed throughout our discussion that the effective
potential term $c_2 \tilde m^2 \Lambda^2$, which appears as a
quadratic divergences in the low-energy theory, dominates and fixes
the dynamics which aligns the soft masses. Such quadratic
divergences are ubiquitous even in supersymmetric theories, when the
supersymmetry is softly broken: while scalar masses are protected
from quadratic divergences, the vacuum energy is not. Could $c_2$
vanish in a particular theory?  Without a symmetry argument,
assuming $c_2 = 0$ is akin to assuming the Higgs is light in a
non-supersymmetric theory. Nevertheless, there have been studies
where the vanishing of $c_2$ was invoked in order to proceed to a
dynamical determination of the effective low energy parameters
\cite{ref:kpz,ref:kprz}, and in particular of the gravitino mass
itself. (Notice that not only $c_2$ but also any terms $c_2'$
arising from intermediate scales must vanish, presumably by the same
mechanism.) While the implications of this idea are interesting, it
is
not clear yet how to implement it in an explicit field-theoretic
model. As a matter of fact, to date the only available examples
\cite{ref:fkz} satisfy $c_2 = 0$ only at 1-loop order (see
\cite{ref:bpr} for an explicit example of its violation at 2-loop
order). Moreover, if such an implementation were found, making
$V_{\rm eff}={\cal O}(\tilde m^4)$, there would still be two
contributions:
one from the MSSM modes, and one which remains as a boundary term
from matching the low- and high-energy theories. While the first
contribution is determined to lowest order by the 1-loop RG
evolution of the Higgs mass and of the cosmological constant, and is
$\sim \tilde m^4 \log (\Lambda/\tilde m)/(4 \pi)^2$, the second
contribution $\sim \xi_0 \tilde m^4$ is in principle unknown. The
first contribution yields a phenomenology similar to the one we have
studied throughout most of this paper. The second can only be
controlled by making our minimality assumption (or some equivalently
strong assumption)---and then, again, similar predictions would be
made. We do not know the relative size of the two contributions. If
we were to treat $\xi_0$ as a usual threshold correction arising
from 1-loop field-theoretic diagrams, we would expect it to be
$\xi_0 \sim 1/16\pi^2$ and hence subdominant in the limit
$\log (\Lambda/\tilde m)\gg 1$, thus weakening  the dependence on
unknown cut-off physics. On the other hand, the MSSM contributions
are suppressed by small Yukawa couplings. In the end, we must plead
at least as much ignorance about the $c_2 = 0$ case as about the
$c_2 \not = 0$ case, and so the minimality assumption is
unavoidable.

Finally, we comment briefly on CP violation. In Ref.~\cite{ref:dt}
the issue of dynamical CP phases was discussed in the
context of the MSSM with universal soft terms. It was shown that,
when the phases of $A$, $M_{1/2}$, $\mu$ and $B\mu$ are promoted to
dynamical variables, the only CP-violating effects have
a CKM origin, {\it i.e.} arise from the Jarlskog invariant $J$, and
thus are suppressed. This was under the strong assumption that no
new sources of explicit CP violation are present. The same
conclusion can be reached in the non-universal case discussed here,
under the parallel assumption that the coefficients in the effective
potential are real. To understand this observation, consider the
limit in which the quark and lepton Jarlskog invariants $J_{q,\ell}$
vanish, and choose
a flavor basis in which the Yukawa matrices are real: in this basis
also
the soft terms relax to real matrices. In Ref.~\cite{ref:dt}, the
invariant
$J$ enters the effective potential at higher-loop order, so its
effect
on the CP-violating phases is further suppressed. In contrast, in a
non-universal scenario, CP violation is already present in
Eqs.~(\ref{eq:mlalign}) and (\ref{eq:aealign}), and
has no further loop suppressions. But as long as $J_\ell$ is not
much larger than its quark counterpart $J_q$, no additional loop
suppressions are needed to ensure that CP-violating quantities such
as electric dipole moments are sufficiently small.

We would like to acknowledge useful and stimulating discussions of
various aspects of this work with M.~Cooper, S.~Dimopoulos, Y.~Nir,
S.~Thomas and F.~Zwirner.

\newpage

\section*{Figure Captions}
\begin{description}
\item[Fig.~1:] Contours of constant branching ratio for $\mu\to
e\gamma$ as functions of the $\mu$ parameter and the unified gaugino
mass $M_{1/2}$ (approximately the wino mass) for various values of
the slepton mass $\tilde m_L$, $\tan\beta$ and the $A$ parameter (as
defined in the text). The black, dark gray, light gray, and white
regions indicate $\br/\br_{\rm exp} < 0.1$, $0.1 < \br/\br_{\rm exp}
< 1$, $1< \br/\br_{\rm exp} < 10$, and $10 < \br/\br_{\rm exp}$,
respectively. The hatched regions are those excluded by LEP I lower
bounds on the mass of the lightest chargino and neutralino.
\end{description}


\begin{thebibliography}{99}
\bibitem{ref:dgt} S. Dimopoulos, G.F. Giudice and N. Tetradis,
Nucl.\ Phys.\ {\bf B454}, 59 (1995).
\bibitem{ref:hor} Y. Nir and N. Seiberg, Phys.\ Lett.\ B\ {\bf 309},
337 (1993); A. Pomarol and D. Tommasini, CERN-TH-95-207,
hep-ph/9507462; L.J. Hall and H. Murayama, UCB-PTH-95/29,
hep-ph/9508296.
\bibitem{ref:msw} L. Wolfenstein, Phys.\ Rev.\ D\ {\bf 17}, 2369
(1978); P. Mikheyev and A. Smirnov, Nuovo Cimento {\bf 9C}, 17
(1986).
\bibitem{ref:neutmass} For a recent analysis of the favored MSW
ranges from the various experiments, see N. Hata and P. Langacker,
Phys.\ Rev.\ D\ {\bf 50}, 632 (1994).
\bibitem{ref:cobe} G.F. Smoot {\it et al.,} Astrophys. J. Lett.,
{\bf 396} L1 (1992).
\bibitem{ref:fluct} S.J. Maddox, G. Efstathiou and W. J. Sutherland,
Mon. Not. R. Astron. Soc. {\bf 246}, 433 (1990); M. Davis, G.
Efstathiou, C.S. Frenk and S.D.M. White, Nature {\bf 356}, 489
(1992).
\bibitem{ref:hot} A. Klypin, J. Holzman, J. Primack and
E. Reg\"os, Astrophys. J. {\bf 416}, 1 (1993).
\bibitem{ref:bm} F. Borzumati and A. Masiero, Phys.\ Rev.\ Lett.\
{\bf 57}, 961 (1986).
\bibitem{ref:hmty} J. Hisano, T. Moroi, K. Tobe and M. Yamaguchi,
KEK-TH-450, LBL-37816, hep-ph/9510309.
\bibitem{ref:sutter} D. Sutter, Ph.D. Thesis, Stanford University,
1995.
\bibitem{ref:us}  R. Rattazzi and U. Sarid, SU-ITP-94-16, RU-95-13,
hep-ph/9505428, Phys.\ Rev.\ D\, in press.
\bibitem{ref:cooper} M. Cooper, (MEGA Collab.), talk at ``Signals of
Unified Theories'', Gran Sasso Nat. Lab., Sept. 1995.
\bibitem{ref:dersav} J.M.~Fr\`ere, D.R.T.~Jones and S.~Raby, Nucl.\
Phys.\ {\bf B222}, 11 (1983); J.-P.~Derendinger and C.A.~Savoy,
Nucl.\ Phys.\ {\bf B237}, 307 (1984).
\bibitem{ref:bh} R. Barbieri and L.J. Hall, Phys.\ Lett.\  B\ {\bf
338}, 212 (1994); R. Barbieri, L.J. Hall and A. Strumia, Nucl.\
Phys.\ {\bf B445}, 219 (1995).
\bibitem{ref:kpz} C. Kounnas, F. Zwirner and I. Pavel, Phys.\ Lett.\
B\ {\bf 335}, 430 (1994).
\bibitem{ref:kprz} C. Kounnas, I. Pavel, G. Ridolfi and F. Zwirner,
CERN-TH-95-11, hep-ph/9502318.
\bibitem{ref:fkz} S. Ferrara, C. Kounnas and F. Zwirner, Nucl.\
Phys.\
{\bf B429}, 589 (1994), ERRATUM-{\it ibid.} {\bf B433}, 255 (1995).
\bibitem{ref:bpr} J. Bagger, E. Poppitz and L. Randall, EFI-95-21,
May 1995, hep-ph/9505244.
\bibitem{ref:dt} S. Dimopoulos and S. Thomas, SLAC-PUB-95-7010.


\end{thebibliography}
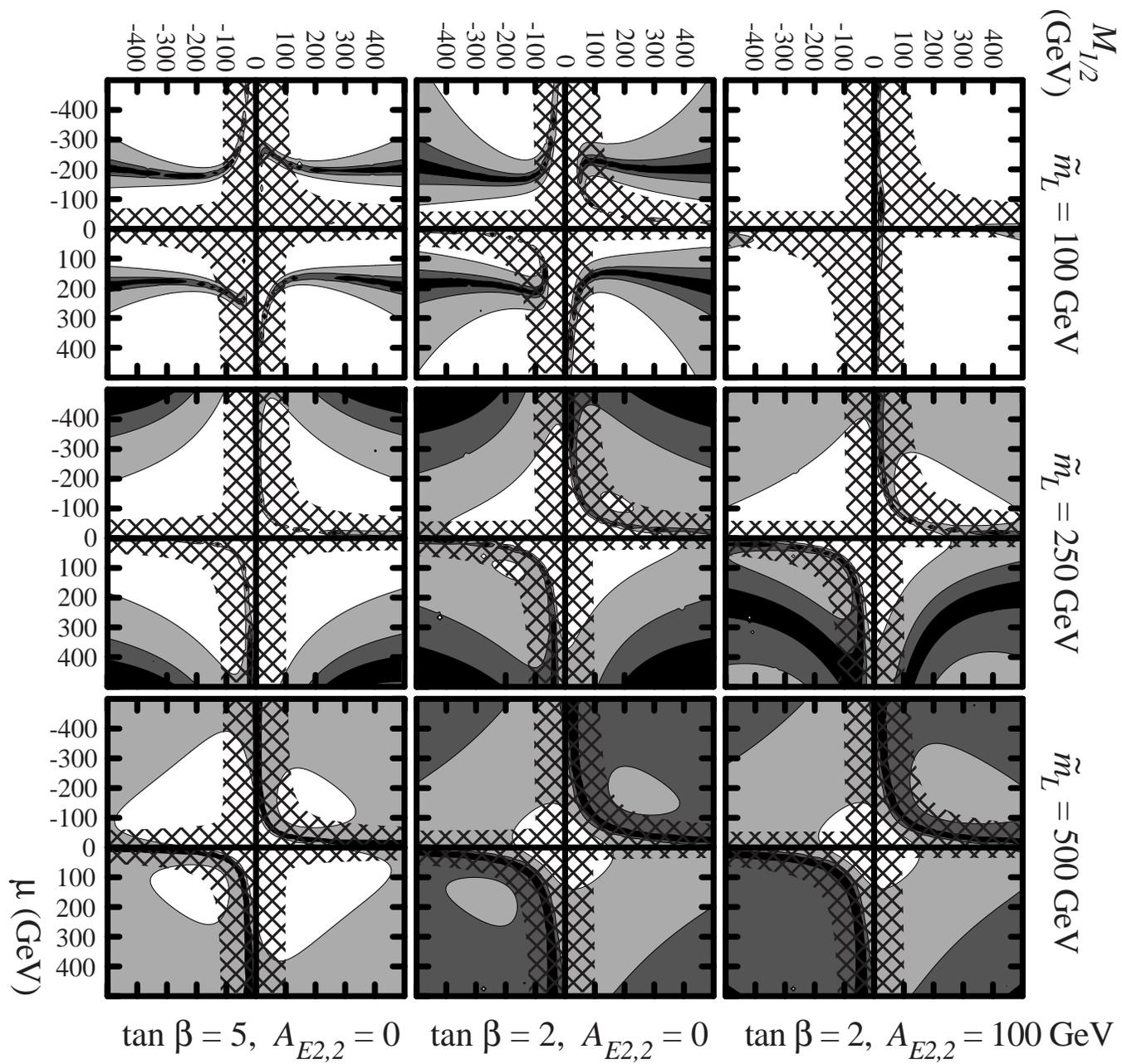
\end{document}